\documentclass[aps,prd,twocolumn,amsmath,amssymb,showpacs,floatfix,nofootinbib]{revtex4-1}

\usepackage{graphicx}
\usepackage{dcolumn}
\usepackage{xcolor}
\usepackage{bm}
\usepackage{natbib}
\usepackage{calligra}
\usepackage[T1]{fontenc}
\usepackage{egothic}
\usepackage[T1]{fontenc}
\newfont{\rsfsten}{rsfs10 scaled 1200}
\newfont{\rsfsseven}{rsfs10 scaled 1200}
\newfont{\rsfsfive}{rsfs10 scaled 1200}
\usepackage{epsfig}
\usepackage{units}
\usepackage[utf8]{inputenc}
\usepackage{multirow}

\newcommand{\aap}{{Astron.~Astrophys.}}

\newcommand{\mnras}{{Mon.~Not.~R.~Astron.~Soc.}}

\advance \voffset by -0.5cm\relax

\def\lsim{\mathrel{\raise.3ex\hbox{$<$\kern-.75em\lower1ex\hbox{$\sim$}}}}
\def\gsim{\mathrel{\raise.3ex\hbox{$>$\kern-.75em\lower1ex\hbox{$\sim$}}}}

\begin{document}

\title{The Black Hole Mass Function from Gravitational Wave Measurements}

\author{Ely D. Kovetz, Ilias Cholis, Patrick C. Breysse and Marc Kamionkowski}

\affiliation{Department of Physics and Astronomy, Johns Hopkins University, Baltimore, MD 21218 USA}

\begin{abstract}
We examine how future gravitational-wave measurements from merging black holes (BHs) can be used to infer the shape of the black-hole mass function, with important implications for the study of star formation and evolution and the properties of binary BHs. We model the mass function as a power law, inherited from the stellar initial mass function, and introduce lower and upper mass cutoff parameterizations in order to probe the minimum and maximum BH masses allowed by stellar evolution, respectively. We initially focus on the heavier BH in each binary, to minimize model dependence. Taking into account the experimental noise, the mass measurement errors and the uncertainty in the redshift-dependence of the merger rate, we show that the mass function parameters, as well as the total rate of merger events, can be measured to $<\!10\%$ accuracy within a few years of advanced LIGO observations at its design sensitivity. This can be used to address important open questions such as the upper limit on the stellar mass which allows for BH formation and to confirm or refute the currently observed mass gap between neutron stars and BHs. In order to glean information on the progenitors of the merging BH binaries, we then advocate the study of the two-dimensional mass distribution to constrain parameters that describe the two-body system, such as the mass ratio between the two BHs, in addition to the merger rate and mass function parameters. We argue that several years of data collection can efficiently probe models of binary formation, and show, as an example, that the hypothesis that some gravitational-wave events may involve primordial black holes can be tested. Finally, we point out that in order to maximize the constraining power of the data, it may be worthwhile to lower the signal-to-noise threshold imposed on each candidate event and amass a larger statistical ensemble of BH mergers.
\end{abstract}


\maketitle

\section{Introduction}

Black holes (BHs) were first identified as a solution to Einstein's field equations
by Schwarzschild in 1916 \cite{Schwarzschild:1916uq}. As early as 1939 it was 
demonstrated that, in principle, they can be formed by the collapse of stars 
\cite{Oppenheimer:1939ue,May:1966zz}. Many decades later, numerous
advances have been made in the study of the physics of black hole formation 
from stars---either by direct collapse or through fallback from supernova explosions
---and core-collapse simulations have been developed to include more and more 
of the relevant mechanisms, most notably the delayed neutrino-driven explosion
mechanism \cite{1990RvMP...62..801B, Janka:2012wk}. 
However, the theory still lacks a clear prediction for the 
number and mass distribution of stellar-mass BHs in the Universe \cite{Fryer:1999mi, 
Fryer:1999ht, O'Connor:2010tk, TheLIGOScientific:2016htt, Fryer:2011cx}.

Observationally, evidence for the existence of stellar-mass BHs in nature has only
recently started to accumulate, thanks to indirect observations of X-ray emission
from accretion of matter from their binary star companions \cite{Motch:1996gw, 
in'tZand:2000zz, Grimm:2001vd, Lutovinov:2004wi, Corbet:2007vn, Russell:2013jva}. 
To date, less than two dozen stellar-mass BHs have been detected in this way, 
most of them in the Milky Way, with a handful of extragalactic candidates 
\cite{Corral-Santana:2015fud, Bogomazov:2016cei}. These X-ray observations, 
however, are limited in reach, becoming more biased as the 
distance from Earth grows, and do not allow a meaningful sample to be gathered 
in order to test the black hole mass function (BHMF) on cosmological scales.

Unfortunately, as their name suggests, black holes do not emit electromagnetic 
radiation and cannot be directly seen. They can be directly {\it heard}, however, 
through gravitational wave (GW) emission from their interaction with binary 
companions such as other black holes and neutron stars \cite{Peters:1963ux, 
Peters:1964zz, Blanchet:2002av, Berti:2009kk, Ajith:2011ec} (or  
from close fly-by encounters, such as tidal disruption events 
\cite{Frank:1976uy, Bloom:2011xk, Stone:2012uk, Ali-Haimoud:2015bfg}). 
The associated time-varying mass quadrupole moment of the two-body 
system results in the emission of GWs, as predicted by Einstein \cite{Einstein:1918btx}. 
These (weak) GW signals have long been sought after in dedicated experiments, 
culminating in the announcement of the first discovery of a coalescing black hole 
binary earlier this year by the LIGO observatory \cite{Abbott:2016blz}. 

With this and subsequent detections of additional binary black hole (BBH) 
mergers \cite{Abbott:2016nmj, TheLIGOScientific:2016pea}, the era of 
gravitational wave astronomy has now finally begun. Overall, advanced 
LIGO in its 2015 O1 run observed three\footnote{The event denoted by 
LVT151012 has a $1.7\sigma$ significance, with an estimated $87\%$ Bayesian 
probability to have been a BBH merger.} 
BH coalescence events, adding nine additional measured black hole masses to 
the current data (the six pre-merger masses range from roughly $7$ to $36\,M_\odot$). 
Over the next decade, improvements in detector sensitivities are expected 
to usher in a wave of newly detected events. LIGO itself is scheduled to 
perform two more runs (O2,3) with increasing sensitivity before commencing 
a multi-year run at its design sensitivity at the turn of the decade. 
Meanwhile, VIRGO \cite{TheVirgo:2014hva} is scheduled to start observing 
during 2017, and plans exist for additional detectors to be built in Japan 
\cite{Aso:2013eba, KAGRA} and India \cite{LIGOIndia} as well. 
These experiments will lead to the discovery of many hundreds of merger 
events per year, providing a rich dataset of black hole statistics to investigate.

In this paper, we demonstrate that the detection of gravitational waves from 
thousands of black hole merger events over the next decade will transform 
our knowledge of the black hole mass distribution, which in turn will shed new 
light on the study of stellar evolution (and termination), star formation history, 
and the progenitor models of binary black holes. 
We show, for instance, that the relation between the slope of the BHMF and the stellar 
initial mass function (IMF) can be probed with high precision. We pay particular attention to
the measurement of the tails of the mass distribution. This can weigh in on 
pressing issues such as the empirical hints of a mass gap between neutron stars 
and black holes \cite{Ozel:2010su, Farr:2010tu} and the abundance of the most massive stellar black holes, 
which is limited by processes such as wind-driven mass loss \cite{Belczynski:2011bn}, preventing 
the heaviest stars from retaining their masses until they collapse to form black holes. 
We then demonstrate that characterizing the distribution of the mass ratio between 
the BBH constituents can be used to probe the efficiency and counterbalance of 
different binary formation mechanisms showing that certain models, such as 
primordial black holes making up the dark matter in our Universe \cite{Bird:2016dcv}, 
can be significantly constrained and possibly ruled out with several years of observation.

Our paper is constructed as follows: 
In Section \ref{sec:Model} we describe our assumptions for the BH mass function 
and its relation to several types of BBH models. Section \ref{sec:Analysis} 
provides the details of our analysis, including details on the choice of parameters 
and noise curves for the future experiments that we consider. Our results are 
presented and explained in Section \ref{sec:Results}. Various subtleties and suggestions for future 
work are discussed in Section \ref{sec:Discussion}. We conclude in Section \ref{sec:Conclusions}.

\section{Modeling the BH Mass Function}
\label{sec:Model}
 
\subsection{The stellar initial mass function}

In this work our reference assumption is that all BHs with mass less
than $\sim100\,M_{\odot}$ originate from the demise of
massive stars (i.e.\, unless stated otherwise, we neglect previous mergers, 
primordial black holes, etc.). To this day, only 29 such BHs have been detected. 
These include 23 BHs discovered through X-rays, 18 of which are galactic  
(see \cite{Corral-Santana:2013uua, Corral-Santana:2015fud, Casares:2013tpa} 
and references therein) and five extragalactic \cite{Orosz:2008kk, 
Steiner:2014zha, Val-Baker:2016veo, Orosz:2007ng, Silverman:2008ss, 
Bogomazov:2016cei}. The remaining ones were recently detected through 
the gravitational waves released from the merger of BBHs with masses 
between $7$ to $36\,M_\odot$ into even more massive end products 
(in total six pre-merger BHs in three merger events) \cite{Abbott:2016blz, 
TheLIGOScientific:2016pea, Abbott:2016nmj}. In addition, there are 42 X-ray 
transients within the Milky Way that are candidates for hosting BHs 
\cite{Corral-Santana:2015fud}. 

Given the currently limited data, in order to model 
the BHMF, we choose to use the better constrained initial mass 
function (IMF) for stars \cite{Salpeter:1955it, Kroupa:2000iv}. The stellar IMF 
is well-described by a multi-part power-law $P(M) \propto M^{-\alpha}$, with $\alpha$ 
taken to be $2.3 \pm 0.7$ for $M > 1 M_{\odot}$ \cite{Kroupa:2000iv}. 
Since the stellar BHs we are concerned with originate from stars with initial 
mass $\gsim 20 M_{\odot}$, we are only sensitive to the slope in the higher 
mass range. We therefore assume for simplicity that the more massive BHs in each binary, 
whose mass we denote by $M_{1}$, will also follow a power-law mass distribution 
$P_{\rm BH}(M_{1}) \propto M_{1}^{\alpha}$, with the value of $\alpha$ corresponding  
to the relevant mass range for the progenitor stars. To facilitate a comparison 
with the LIGO collaboration results in Ref.~\cite{TheLIGOScientific:2016pea}, we 
follow their assumptions and set $\alpha=2.35$ as our fiducial value.

We will examine the precision with which $\alpha$ can be constrained by advanced 
LIGO observations over the coming decade (assuming a power-law distribution 
remains consistent with the data). After the first observations of the three coalescence 
events GW150914, LVT151012 and GW151226, LIGO has constrained $\alpha$ to 
be $2.5^{+1.5}_{-1.6}$ at 90$\%$ credible interval. We shall see that future 
observations will go well beyond this precision. 

\subsection{Endpoints in the mass function of BHs: the mass gap and mass cap}

The boundary mass between a neutron star (NS) and a BH is expected not 
to exceed $M \simeq 3 M_{\odot}$, with Ref.~\cite{Lattimer:2012nd} suggesting a 
value of 2.5 $M_{\odot}$ as the highest possible neutron star mass. In current 
observations, the most massive well-measured neutron star is $2.01\pm 0.04 M_{\odot}$ 
\cite{Antoniadis:2013pzd} and the least massive BH is $\sim 4.4 M_{\odot}$ \cite{Filippenko:1999zv}.
This has led various authors to suggest the presence of a mass gap between 
approximately 2-5 $M_{\odot}$ \cite{Bailyn:1997xt, Ozel:2010su, Farr:2010tu}. 
It remains to be seen whether this empirical mass gap originates from selection effects 
in the still small sample of BHs (although based on the IMF, we would expect 
lower-mass BHs to be more abundant), or from the underlying assumptions
that enter into their mass estimates \cite{Kreidberg:2012ud}, or whether it is indeed 
suggestive of the properties of the relevant supernovae and their progenitor stars 
\cite{Belczynski:2011bn, Fryer:2011cx, Kochanek:2013yca}. Our results below will 
show that the mass gap, especially if the transition is as sharp as currently indicated 
by both experiment and pertaining theoretical models, can be extremely well 
constrained by gravitational wave measurements.

Meanwhile, we also expect an upper bound on the stellar mass allowing for BH 
formation. Wind-driven mass loss causes stars too massive to lose a significant 
portion of their mass before they evolve to produce black holes \cite{Crowther:2006dd} 
(see though Ref.~\cite{Smith:2014txa}). Stars heavier than 
$\sim 300 M_{\odot}$ have not been observed to date \cite{Crowther:2010cg}.   
Meanwhile, various works (see e.g.\ Refs.~\cite{Komossa:1998zw, Weidner:2003eh, 
2006MNRAS.365..590K, Oey:2005mn, Banerjee:2011gh}) have suggested a 
lower and more conventional upper bound on the stellar mass $\simeq 150 M_{\odot}$). 
We wish to explore the sensitivity of GW measurements to this important quantity. 

We therefore parametrize the BHMF as
\begin{equation}
P(M_1)\propto M_1^{-\alpha}\mathcal{H}(M_1\!-\!M_{\rm gap})e^{-M_1/M_{\rm cap}}
\label{eq:BHMF}
\end{equation}
where $M_1$ is the mass of the heavier binary component, $\alpha$ is a power law 
with a fiducial value of $2.35$ (to match the Kroupa mass function \cite{Kroupa:2000iv}), 
$M_{\rm gap}$ is the NS-BH transition cutoff, which we take to be sharp,  
$M_{\rm cap}$ is a (shallower) exponential upper cutoff on the BH mass and 
$\mathcal{H}$ is the Heaviside function.

\begin{figure}
\begin{centering}
\includegraphics[width=\columnwidth]{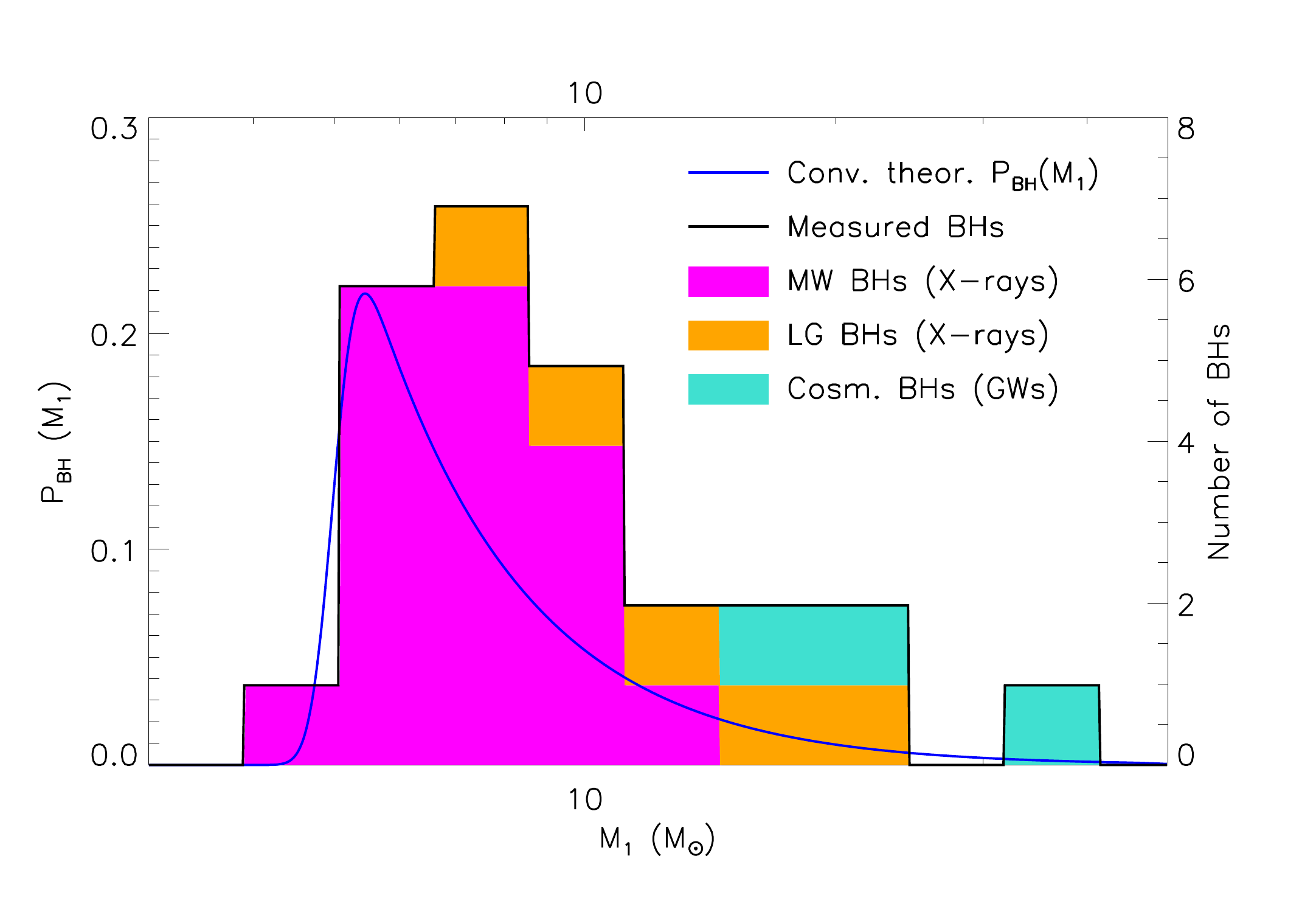}
\end{centering}
\vspace{-0.3in}
\caption{The distribution function $P_{\rm BH}(M_{1})$ of the heavier BH mass $M_{1}$ 
in either BBHs or BH-star binaries. The blue line shows the theoretical prediction 
of Eq.~(\ref{eq:BHMF}) convolved with a Log Normal distribution that has a 5$\%$ error. 
The black histogram describes the observed distribution from 26 BHs. The red 
histogram is similar, but is based solely on the 18 dynamically confirmed BHs 
observed in X-rays that originate in the Milky Way. The yellow histogram 
stacked on the red one includes the five BHs observed through X-rays at LMC, 
IC 10, M33 and NGC 300. The green histogram stacked on the other two includes 
the three $M_{1}$ masses detected by LIGO.}
\label{fig:Pm1Measured}
\end{figure}

In Fig.~\ref{fig:Pm1Measured}, we plot the mass distribution of BHs observed 
in X-ray binaries as well as merging BBHs. For the latter we only show the mass 
of the heaviest of the two BHs $M_{1}$ (before the merger event). We divide 
the X-ray binaries to Galactic (denoted by MW) and extragalactic events that have 
been detected in the Local Group or in its direct vicinity (denoted by LG). For the 
more distant X-ray binaries there is some selection effect towards more massive 
systems. In solid blue we plot the theoretical probability density 
function of $M_{1}$, Eq.~(\ref{eq:BHMF}). To account for reasonable uncertainty 
in the mass estimation, we convolve the BHMF with a log-normal distribution with a 
$5\%$ error in the mass (see more on the treatment of mass measurement errors 
in the next section).  

While the statistics based on GW measurements in this figure are quite modest, 
we expect up to thousands of BBHs to be observed in the next decade. 
These observations will surpass the X-ray sample in size by a significant factor 
and as we demonstrate will dramatically affect the observed mass distribution.

\subsection{Redshift-dependence of BH-BH merger rates}

In addition to learning about the distribution of BH masses, inferring their 
local merger rate and their redshift distribution is of great interest. 
X-ray observations have resulted in the detections of X-ray transients from 
binaries with a BH and a companion star that extend only as far away as 
NGC 300 (just in the vicinity of the Local Group, at a distance of 1.8 Mpc 
\cite{Rizzi:2005wq}), with the majority of transients detected inside the 
Milky Way. Looking forward, with gravitational wave measurements we
will be able to probe merger events of BH binaries at cosmological 
distances up to Gpcs, or in terms of redshift up to $z \sim 0.3$ with the 
current instrumental sensitivity \cite{TheLIGOScientific:2016zmo} and 
up to $z \simeq 1$ with the expected advanced LIGO design \cite{Aasi:2013wya}. 
In the future, experiments such as the Einstein Telescope (ET) may reach 
detection thresholds corresponding to redshifts  $z \gtrsim 10$ \cite{Sathyaprakash:2011bh}. 

After the first three detections, the local rate of BBH mergers has been 
estimated by LIGO to be $53^{+100}_{-40}$ Gpc$^{-3}$yr$^{-1}$, assuming 
that the three events do not follow any specific mass function, and 
$99^{+138}_{-70}$ Gpc$^{-3}$yr$^{-1}$ if their heaviest mass $m_{1}$ follows 
a mass function scaling as $M_{1}^{-2.35}$ down to masses of 5 $M_{\odot}$ 
\cite{TheLIGOScientific:2016pea}. This rate is not expected to be redshift independent 
and instead is sensitive to the metallicity environment in which BBHs form, 
as well as the typical timescale for them to merge. This timescale is referred to 
as the time delay. For stellar BBHs it is of the order of 100s of Myrs to Gyrs 
(see Ref.~\cite{Dominik:2013tma} and references therein). If BBHs originate mainly 
in low metallicity environments, their merger rates $R(z)$ peak at high 
redshifts, while if the time delay is large, that peak moves to lower $z$. 

With the cosmological distances $z \simeq 1$ to be probed in the next decade, 
it is preferable to use a generic parametrization that accounts both for the local 
rate uncertainties and for the uncertainties in the redshift distribution.
In this work we use a simplified parametrization given by
\begin{equation}
R(z) = R_{a} (1+z)^{R_{b}}.
\label{eq:Rmz}
\end{equation}
This monotonic behavior of $R(z)$ cannot be valid up to arbitrarily high 
values of $z$. If most BBH systems form in high metallicity environments, 
then with sufficient LIGO observations we should see a deviation from this 
parametrization, leading to lower best-fit values for $R_{b}$. 
Furthermore, even for very low metallicity environments and ignoring any 
time delays, $R(z)$ is expected to drop for $z > 4$ \cite{TheLIGOScientific:2016htt, 
Dominik:2013tma}. Thus, Eq.~(\ref{eq:Rmz}) can be considered relevant only 
for the BBH coalescence observations from Advanced LIGO. With ET sensitivities, 
for example, such a parametrization will require modifications, to account for 
the decreasing merger rate at high redshifts. 

For this work, we take as fiducial values for $R_{a}$ a local rate of  $99$ 
Gpc$^{-3}$yr$^{-1}$ and for the power-law index $R_{b}$ a value of 2. 
These provide a good fit to the approximations in Ref.~\cite{TheLIGOScientific:2016htt}.
In our analysis of the BHMF below, we will either hold the rate parameters fixed---an optimistic 
assumption---or marginalize over them---which is a conservative assumption, 
appropriate if there is no other source (based on theory or experiment) to provide more information. 
Our constraints on the merger rate parameters themselves are calculated when
marginalizing over the other parameters, to avoid additional assumptions that 
are often made in the analysis (such as in Ref.~\cite{TheLIGOScientific:2016pea}).

\subsection{The mass ratio in BBHs and the 2D BHMF}

The measurement of the mass of the lighter black hole in the merging 
binaries presents both a challenge and an opportunity. On the one hand,
it would double the amount of information that can in principle be used to 
infer the BHMF parameters. On the other hand, this requires  
the two-dimensional distribution of the masses in the binary, which is
strongly model dependent. Binary formation mechanisms vary greatly 
in their prediction for the binary mass ratio $q\equiv M_2/M_1$. 

In the common envelope scenario, where the BHs are formed from binary stars 
which subsequently transform into BHs, with mass exchanged between the two 
throughout their common evolution \cite{Belczynski:2001uc}, we expect the ratio 
to be larger than it would be under the naive assumption that the mass values for 
both $M_{1}$ and $M_{2}$ are drawn randomly from the same distribution 
($\propto M^{-\alpha}$)  \cite{Fryer:1999ht, Woosley:1994wc, Marchant:2016wow}. 
In this scenario, the BH masses depend also on how effective the Wolf-Rayet 
phase is, during which significant mass loss of the progenitor stars takes place. 

Dynamical formation of binaries tends to lead to larger values of $q$
\cite{O'Leary:2016qkt, Rodriguez:2016kxx}. One of the reasons is dynamical friction. 
In globular clusters and in environments where the BHs fall 
towards the center of a potential, it causes the most massive BHs to fall in first, 
and thus the first binaries to form in/close to the center of the potential are the pairs 
that contain the most massive BHs. Then the next most massive stars fall in and 
create companions, etc. Another reason to expect high values of $q$ is scattering 
processes, either involving the BBH and an additional single BH, or between two 
BBHs. Simulations have shown that in dense environments, binaries tend to 
exchange components, preferentially ejecting their smaller partners in favor of more 
massive companions \cite{Sigurdsson:1993zrm, Rodriguez:2016kxx, Rodriguez:2016avt}.

Meanwhile, more exotic mechanisms may lead to more extreme $q$ distributions.
In the primordial black hole scenario, for example, observational constraints
limit the extent of the mass function and if the mass distribution is narrow, we 
expect $q$ to be roughly unity for all PBH binaries. In this case, of course, the 
2D distribution would be very different than for stellar black holes, as both masses
would have similarly narrow distributions. We shall return to this point below.

We therefore describe the 2D BHMF using Eq.~(\ref{eq:BHMF}) and 
\begin{equation}
P(M_2)\propto (M_2/M_1)^{\beta}\mathcal{H}(M_2\!-\!M_{\rm gap})\mathcal{H}(M_1\!-\!M_2),
\label{eq:pm2}
\end{equation}
where $M_2$ is the mass of the lighter binary component and $\beta$ 
is a power law with a value that depends on the BBH progenitor model. 
As a fiducial value we follow the LIGO analysis in their 
Ref.~\cite{TheLIGOScientific:2016pea} and set $\beta=0$, i.e.\ a uniform 
distribution for $M_2$ in the range $[M_{\rm gap},M_1]$.
In order to provide a result that is less progenitor-model dependent, however, 
we differ from the analysis in Ref.~\cite{TheLIGOScientific:2016pea} 
in that rather than {\it fixing} this assumption for the probability distribution of the 
lighter mass in each binary, we first limit the analysis to contain only the number 
counts of the heavier mass in each binary and calculate the constraints on the
BHMF parameters while {\it marginalizing} over the mass ratio parameter introduced 
in Eq.~(\ref{eq:pm2}). As explained below, we then extend our analysis 
to use the full two-dimensional mass distribution and constrain the mass ratio in 
tandem with the BHMF parameters.

\section{Analysis}
\label{sec:Analysis}

\subsection{Experimental signal-to-noise ratio}

The detectability of GWs from a coalescence event depends on the relevant 
signal-to-noise ratio ($S/N$), which for a single interferometer detector is given by
\begin{equation}
({S/N})^{2} = \frac{4}{5} \int_{f_{\rm min}}^{f_{\rm max}}
df \frac{h_{c}^{2}(f)}{S_{n}(f) (2 f)^{2}},
\label{eq:SNR}
\end{equation}
where $h_{c}(f)$ is the observed strain amplitude and $S_{n}(f) = h_{n}^{2}(f)$ 
is the strain noise amplitude. We follow the same assumptions and parametrization  
of the coalescence signal from two merging BHs of Ref.~\cite{Cholis:2016kqi} 
(see references therein), including though only the dominant quadrupole radiation. 
We assume that a given event is ``detected'' when the $S/N$ in a single interferometer 
satisfies
\begin{equation}
S/N>8.0.
\label{crit2}
\end{equation}
This single detector criterion approximately translates into $S/N>12$ for a network 
of the two LIGO detectors and given the overall lower sensitivity of the VIRGO 
interferometer is roughly correct for the combination of the three as well. This criterion 
is conventionally used as the threshold for a GW detector network to be able to identify 
the GW signal from a merging binary (e.g. \cite{Abadie:2010cf}). 

Since a large sample of events is necessary in order to understand the averaged 
properties of BBHs over cosmological distances, we will also consider less 
stringent $S/N$ thresholds for the flagging of candidate GW events as detected mergers.  
The inevitable tradeoff between more statistics and larger individual mass estimation 
errors may motivate future advances in the estimation of the component BH masses 
in the binaries. To decouple our reported forecasts from our noise-related assumptions 
(we neglect the observing duty cycle of the experiment, for example), our main results 
will be presented in terms of the total number of detected coalescence events (for 
bookkeeping purposes, we will quote the corresponding number of observation years 
under our assumptions). Imposing a lower $S/N$ threshold will simply result in attaining 
the same sample size earlier in time (and vice versa).     
The noise power spectrum used in our calculations refers to advanced LIGO in its final 
design sensitivity, for which we adopt a noise model based on the analytical approximation 
of Ref.~\cite{Ajith:2011ec} (see their Eq.~(4.7)) to the official advanced LIGO design 
noise curve \cite{Shoemaker2010}. We set the lower frequency limit at 
$f_{\rm min}=20\, {\rm Hz}$, above which these noise curves match very well.

\subsection{Mass measurement error}

In order to account for errors in the individual masses, we use two methods. 
First, we assume a relative error in the mass measurement and so to get the 
{\it observed} probability distribution function, we convolve Eq.~(\ref{eq:BHMF}) 
with a log-normal distribution
\begin{eqnarray}
P(M_{\rm obs})&=&\iint P(M_{\it th})P_{\rm G}(x)\delta\left(M_{\rm obs}-
xM_{\rm th}\right)dx\,dM_{\rm th}  \cr
&=&\int P(M_{\rm th})P_{\rm G}\left(M_{\rm obs}/M_{\rm th}\right)
dM_{\rm th}/M_{\rm th}
\label{eq:Pmobs}
\end{eqnarray} 
where $M_{\rm th}$ is the real value of the mass (which follows the theoretical 
PDF in Eq.~(\ref{eq:BHMF})), $M_{\rm obs}$ is the observed mass and the 
relation between them is given by $M_{\rm obs}=xM_{\rm th}$, where $x$ is 
distributed normally, $x\sim\mathcal{N}(1,\sigma^2)$, and 
$P_{\rm G}=\frac{1}{\sqrt{2\pi\sigma^2}}e^{-(x-1)^2/2\sigma^2}$.
For the calculations below we take $\sigma=0.05$ (a $5\%$ relative mass error). 
We have verified that our conclusions are not sensitive to this choice, as 
long as the error is not considerably larger (as will be explained below, the parameter 
most sensitive to the measurement uncertainty is the $M_{\rm gap}$ cutoff).

Secondly, in the Fisher analysis below, we use a logarithmically-binned BHMF 
measurement, with bins wide enough to ensure minimal cross-over between 
bins due to measurement errors (and again, the conclusions 
are not sensitive to the particular bin width used here). 

\subsection{The total number of detected merger events}

The observable we consider in this work is the total number of detected merger 
events with a given BH mass ($M_1$, the mass of the heavier BH in the 1D case, 
or the two masses $M_1,M_2$ in the 2D case).
The theoretical prediction for this quantity, based on our model for the BH mass
probability distribution function and the merger rate is given by
\begin{eqnarray}
\frac{dN(M_1)}{dM_1} &=& 4\pi A_{M_1}P({M_1}) \int\limits_{M_{\rm gap}}^{M_1} A_{M_2}P(M_2)dM_2  \cr 
&&\times\int\limits_0^{z_{\rm max}(M_1, M_2)}\frac{c\chi(z)^2R(z)}{(1+z)H(z)}dz,
\label{eq:dN}
\end{eqnarray}
where $A_{M_1}$ and $A_{M_2}$ are the normalizations of the two PDFs in 
Eqs.~(\ref{eq:BHMF}) and (\ref{eq:pm2}); The upper limit on the redshift integral, 
$z_{\rm max}(M_1, M_2)$, is the maximum redshift up to which the merger of a BBH with masses $M_1,M_2$ 
can be detected with the experimental setup considered; $H(z)$ is the Hubble parameter
and $\chi(z)$ is the comoving distance. To incorporate the measurement error,
we use the observed PDFs, Eq.~(\ref{eq:Pmobs}). In the 2D case, we use 
$dN(M_1,M_2)/dM_1dM_2$, defined similarly to Eq.~(\ref{eq:dN}),  only
dropping the first integration.

\subsection{Fisher matrix constraints}

We will use the Fisher matrix formalism to study how well the BHMF parameters, 
the merger rate and the binary mass ratio can be constrained using GW measurements. 
This method assumes that the likelihood distribution of the parameter values is a 
multivariate Gaussian, centered on chosen fiducial values. The Fisher matrix 
$F_{\mu\nu}$ for a model with parameters $p_\mu$ is given by \cite{Jungman:1995av,Jungman:1995bz}
\begin{equation}
F_{\mu\nu}=\sum_i\frac{1}{\sigma_i^2}\frac{\partial N_i}{\partial p_\mu}\frac{\partial N_i}{\partial p_\nu},
\label{Fisher}
\end{equation}
where $N_i$ is the number of events in each mass bin $i$
\begin{equation}
N_i = \int_{M_{\textrm{min},i}}^{M_{\textrm{max},i}}\frac{dN(M_1)}{dM_1}dM_1
\end{equation}
is the number of detected BHs in a mass bin with edges $[M_{\textrm{min},i},M_{\textrm{max},i}]$.
In our analysis below we divide $N(M_1)$ into $30$ logarithmic bins from $M_1=4$ to $M_1=120$. 
We assume the bins
obey Poisson statistics and take $\sigma_i^2=N_i$ as the expected variance in $N_i$. 
As long as the bins are wide enough and we have ample statistics, this should be 
a reasonable assumption. The Fisher matrix computed from Eq.~(\ref{Fisher}) 
can then be inverted to obtain the covariance matrix of the model parameters.

We compute the Fisher matrix using the six parameters introduced earlier: 
the three BHMF parameters, $\left\{\alpha, M_{\rm gap}, M_{\rm cap}\right\}$, 
the two merger rate parameters, $\left\{R_a, R_b\right\}$, and the power-law 
index $\beta$ of the mass ratio distribution. Our fiducial choice of parameter values
is $\alpha = 2.35$, $M_{\textrm{gap}} = 5\,M_{\odot}$, $M_{\textrm{cap}} = 60\,M_{\odot}$, 
$R_{a} = 99$ Gpc$^{-3}$yr$^{-1}$, $R_{b}=2$ and $\beta = 0$. As we explain 
below, depending on the desired forecast, in some cases we focus on certain 
parameters and marginalize over the value of others. The cosmological parameters
that enter our calculations are taken from Ref.~\cite{Ade:2015xua}. 
Finally, we note that when addressing the 2D mass distribution, the formalism is 
almost identical. We simply replace $N(M_1)$ with $N(M_1,M_2)$, divide into 
$15\times15$ bins and $N_i$ is then the number of BHs in each 2D mass bin.

\section{Results}
\label{sec:Results}

\begin{figure}
\begin{centering}
\includegraphics[width=0.95\columnwidth]{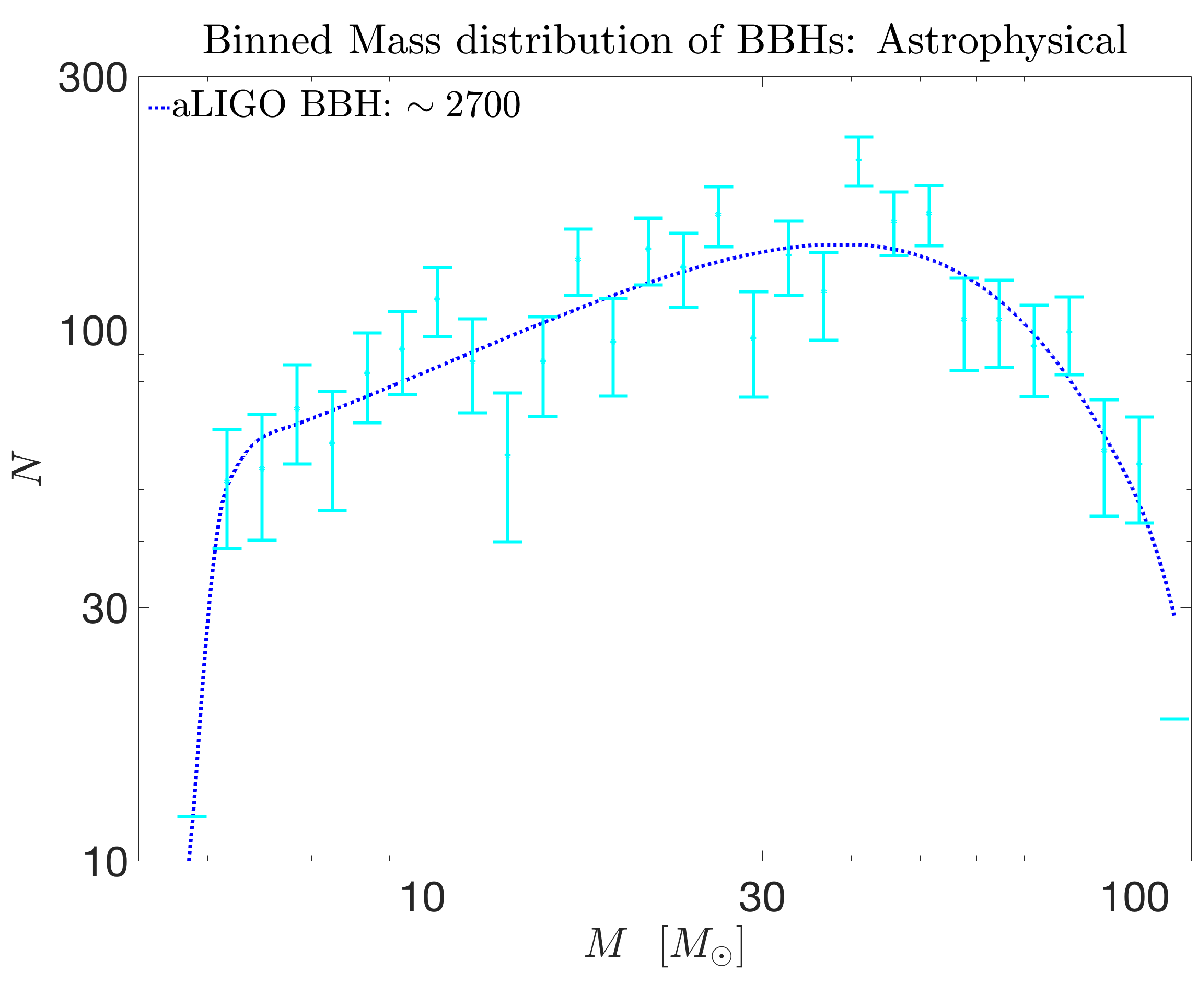}
\end{centering}
\caption{The logarithmically-binned distribution of the mass of the heavier 
component in BBH mergers with $2700$ BHs (as predicted for 6 years of aLIGO at final design 
sensitivity given our noise assumptions). The dashed blue line shows the theoretical 
expectation with our set of fiducial parameters, and the cyan error bars indicate the expected
distribution assuming $\sqrt{N_i}$ Poisson noise in each mass bin.}
\label{fig:1Ddist}
\end{figure}

Our first result is a calculation of the number of observed events as a function 
of the mass of the heavier member of the BBH, given in Eq.~(\ref{eq:dN}). 
In Fig.~\ref{fig:1Ddist} we plot the predicted mass distribution of observed 
BBH mergers for six years of advanced LIGO observations (at design sensitivity), 
which totals $\sim\!2700$ events for our choice of parameters. 
As we can see, this function has a peak at masses much heavier than the peak of 
the PDF in Fig.~(\ref{eq:BHMF}). This peak is in fact very close to the mass of the 
heavier BH in the first event detected by LIGO, $M_1\sim36\,M_\odot$. 
This stems from the fact that heavier masses yield mergers with larger GW strain 
amplitudes, and can therefore be detected at  greater luminosity distances. 
The resulting increase in detectable {\it volume} is more than enough to compensate 
for the negative mass function slope $-\alpha=-2.35$. 

The number of expected GW events depends on the choice of both the lower mass 
cutoff $M_{\rm gap}$ and the upper cutoff $M_{\rm cap}$. It also 
naturally depends on the assumed rate of merger events throughout the observable 
redshift volume\footnote{We note that the fiducial rate we use, calculated in 
Ref.~\cite{TheLIGOScientific:2016pea}, relies on specific assumptions regarding the 
BHMF parameters. Thus if the true values of the latter deviate significantly from the 
ones assumed here, then the rate of 99 Gpc$^{-3}$yr$^{-1}$ would also have to be 
replaced. We neglect this subtlety in our analysis.}. In our results below we investigate 
the degeneracies between the BHMF, the coalescence event rate and binary mass ratio parameters. 

\begin{figure}
\begin{centering}
\includegraphics[width=0.95\columnwidth]{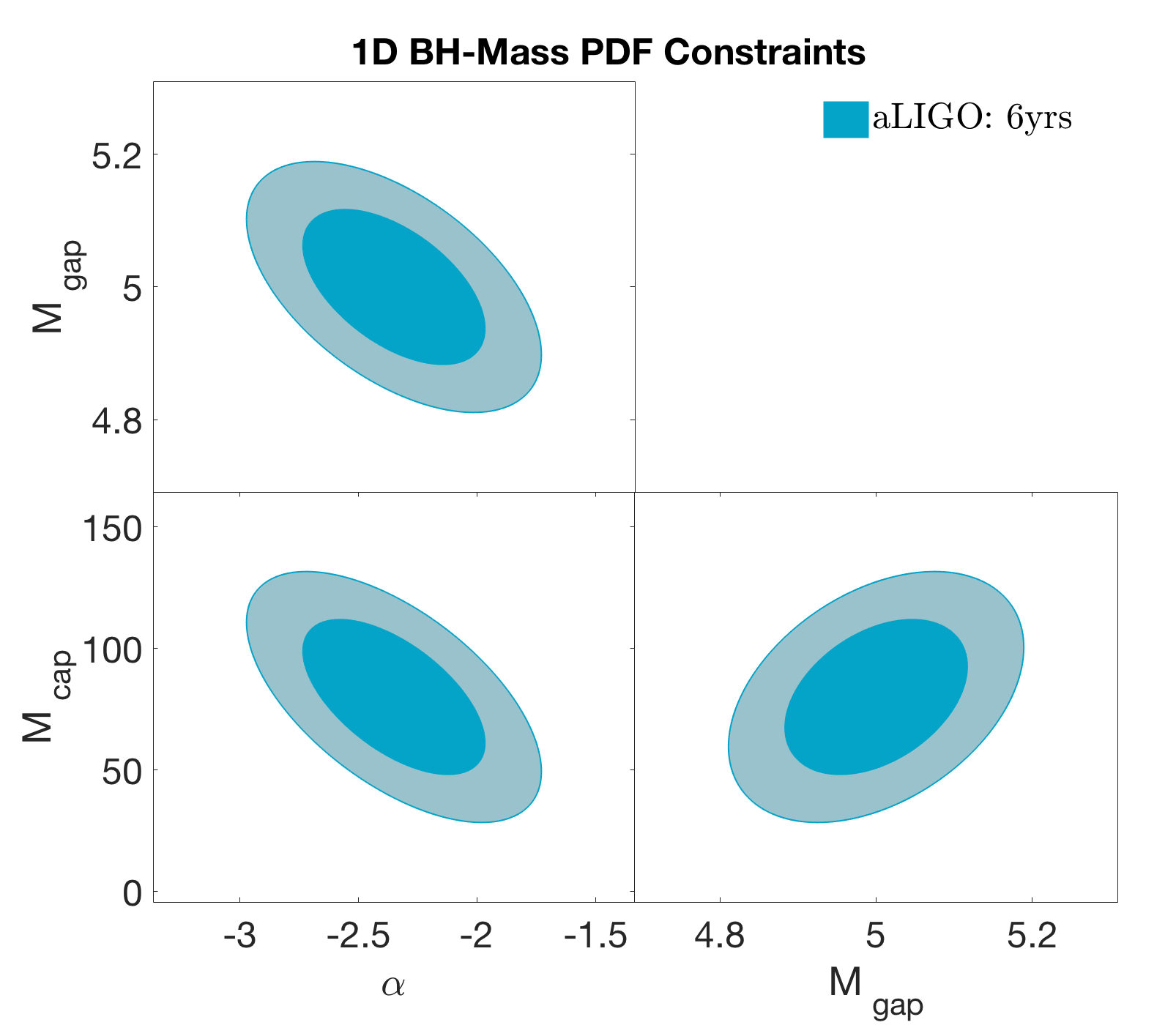}
\end{centering}
\caption{Constraints on the BHMF parameters with 2,700 BHs (6 years of aLIGO observations).
We marginalize over the merger rate parameters and over the value of the mass ratio power-law.
The resulting constraints are promising. The constraint on $\alpha$ is tighter than the current best
constraint on the IMF power law, while the lower mass cutoff can be measured well enough to 
confidently confirm or rule out a NS-BH mass gap.}
\label{fig:1Dparams}
\end{figure}

Employing the Fisher analysis described above, we use this prediction for the observed 
mass distribution of the heavier black hole to calculate the resulting constraints on 
the three BHMF parameters. These are shown in Fig.~\ref{fig:1Dparams}.
Here we marginalize over the merger rate parameters (we return to the merger rate below).
From Fig.~\ref{fig:1Dparams}, we see that the detection of $\sim\!2700$ events 
will yield constraints on the BHMF parameters ranging from $2\%$ to 
$40\%$ (at 1-$\sigma$). The excellent sensitivity to the mass gap $M_{\rm gap}$ 
is such that if the true minimum mass of stellar black holes is indeed $\sim5\,M_\odot$, 
then we can reject the hypothesis that the distribution extends all the way down to the 
upper limit on neutron star masses of $\simeq 2\,M_\odot$, with high significance ($\gg\!5\sigma$). 
The sensitivity to the mass gap depends, however, on both the value of the cutoff and 
the assumed measurement error. If the cutoff is $M_{\rm gap}=4$, we will only be able to 
confirm the gap at the $\gtrsim3\sigma$ level, unless the measurement error can be reduced to $<5\%$ 
(and if the cutoff is even lower, it may remain undetectable until a more sensitive future experiment 
provides much more statistics). A $\gtrsim3\sigma$ level of confidence will also correspond to a case
where $M_{\rm gap}=5$ and the measurement error is $>10\%$ in this mass range.

The BHMF power-law $\alpha$ is constrained to roughly $15\%$, which will suffice 
to detect considerable deviations from the power-law index of the IMF that could 
hint at possible selection effects in the black hole formation mechanism.
The constraints on the top end of the distribution are weaker still, governed by the 
tradeoff between the decreasing probability to see heavier masses versus the increase 
in signal to noise of their merger events. This weaker constraint is also a result of the 
fact that our model uses an exponential upper cutoff, which decays fairly slowly 
compared to sharper cutoffs, like we use for the lower mass end.

It is important to reemphasize that in the calculations above we also marginalized 
over the mass ratio between the two masses in each binary (the mass of the second 
black hole in each merger enters the calculation above indirectly as it affects its detectability). 
Therefore, as we use only the heavier mass measurements, we are discarding half(!) 
of the data at our disposal. However, as the mass of the lighter black hole in the 
binary tends to be highly dependent on its counterpart in many of the progenitor 
models, this ensures that the BHMF constraints we achieve are less model-dependent.

In the second part of this section we turn our focus to the two dimensional distribution, 
and show that in fact one could use the complete dataset, i.e.\ the two-dimensional mass 
distribution, precisely to distinguish between different progenitor models, as well as 
improving the constraints overall.

\begin{figure}[h!]
\begin{centering}
\includegraphics[width=0.95\columnwidth]{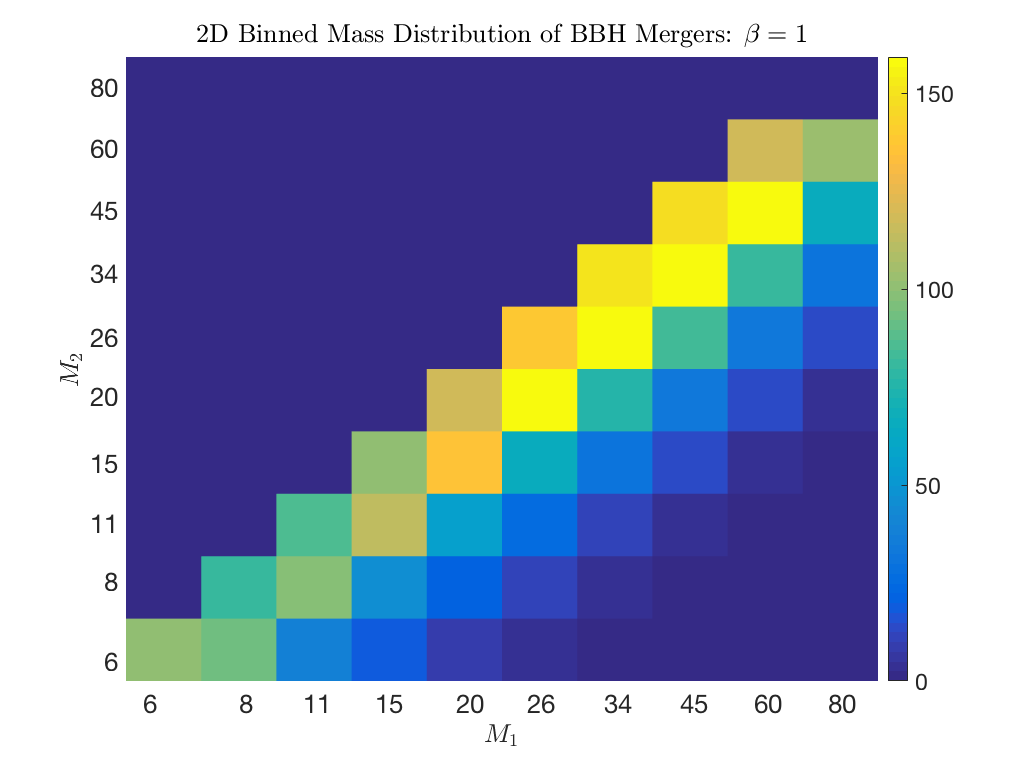}
\includegraphics[width=0.95\columnwidth]{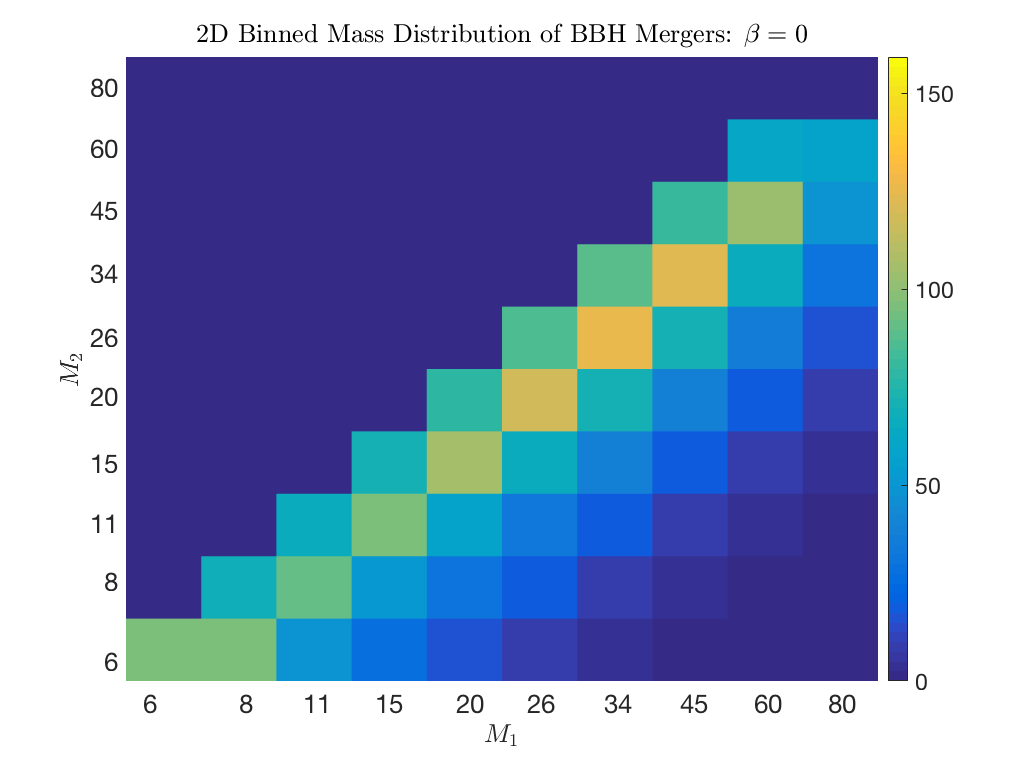}
\includegraphics[width=0.95\columnwidth]{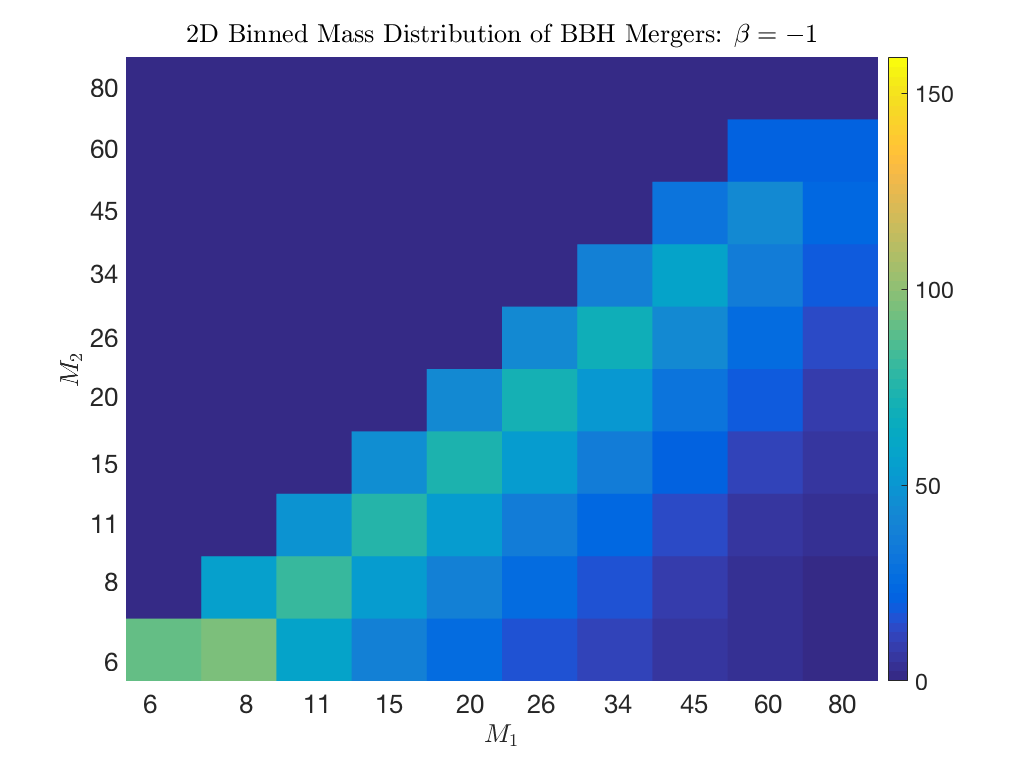}
\end{centering}
\caption{The 2D mass distribution of BBH mergers, for three different mass-ratio power laws.
While the projection of these plots down to one mass dimension would look qualitatively similar 
to Fig.~\ref{fig:1Ddist}, it is clear that the 2D mass distribution contains more information.  
This can be used to break degeneracies between parameters, as demonstrated in our subsequent results.}
\label{fig:2Ddists}
\end{figure}

In Fig.~\ref{fig:2Ddists} we plot the 2D mass distribution of BBH mergers, comparing 
the results for $\beta=-1,0,1$, which would correspond qualitatively to different 
(and distinct) progenitor models, as explained in the previous Section. 
Comparing the number counts along the diagonal and in the bottom corner of these plots, 
it is evident that information lost in the projection to the 1D $M_1$-analysis  
can be used to improve the sensitivity to the mass ratio.

\begin{figure*}
\begin{centering}
\includegraphics[width=\columnwidth]{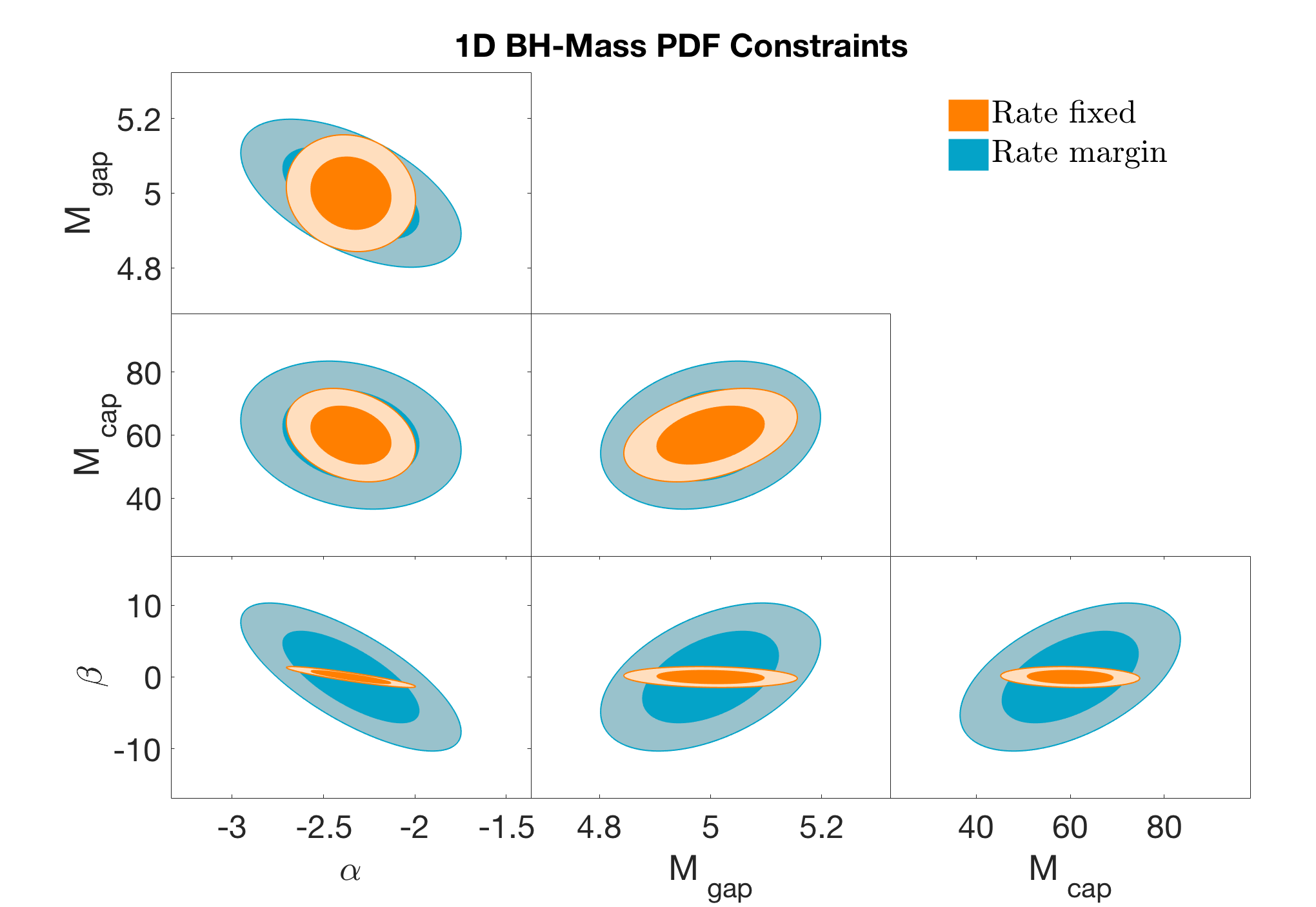}
\includegraphics[width=\columnwidth]{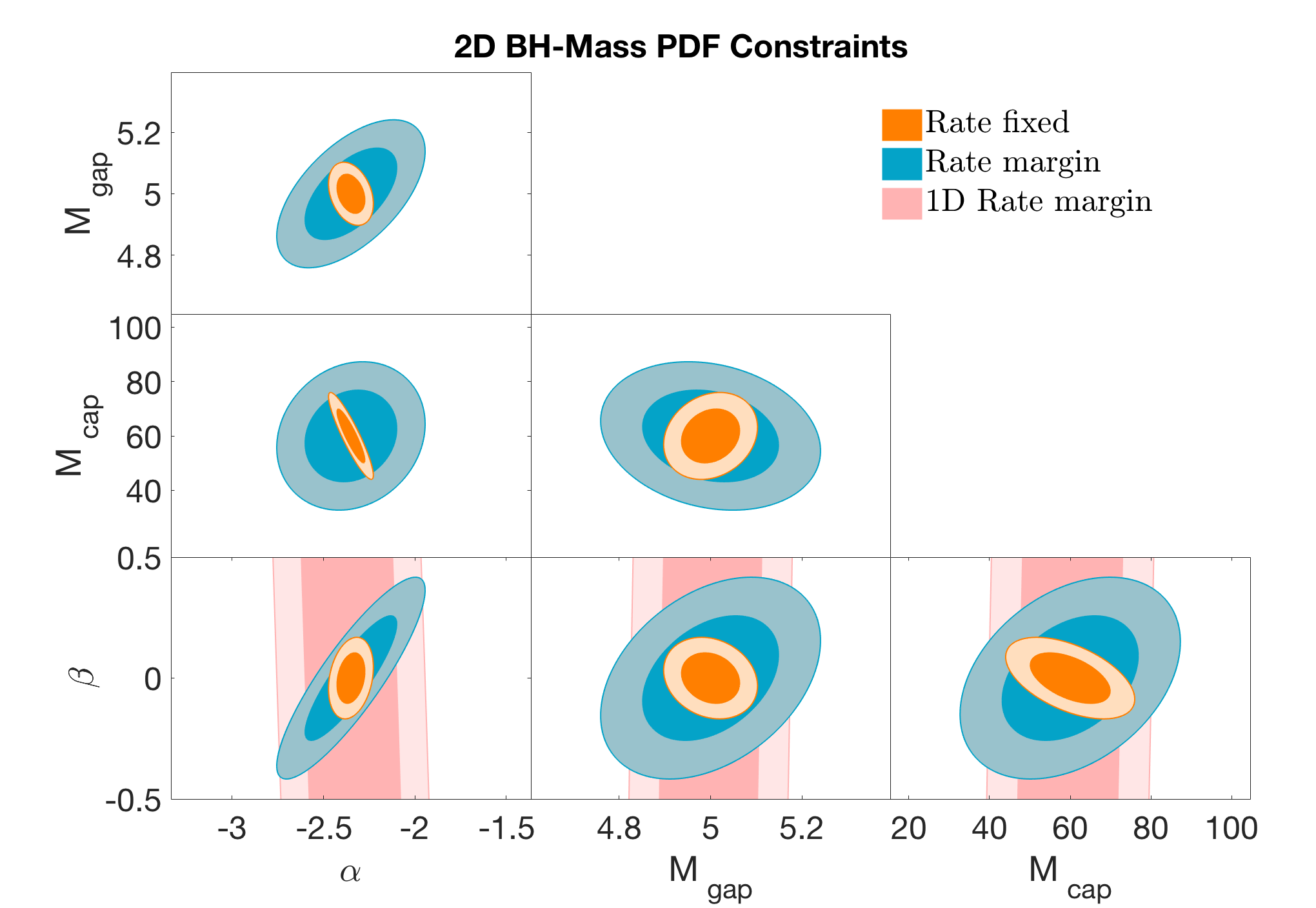}
\end{centering}
\caption{Constraints on the BHMF parameters and the mass-ratio power law, using 1D 
and 2D data. While for measuring the BHMF parameters the improvement in going from 
one to two dimensions is modest, there is a stark difference in the constraints on the 
mass ratio parameter, which tighten by a factor $\sim20$. Note that the pink ellipses 
in the bottom right-hand plots (where the y-axis scale for $\beta$ is much smaller) seem 
narrow, yet they do extend to a width similar to the blue ones when zooming out.}
\label{fig:2Dvs1Dparams}
\end{figure*}

\begin{figure}
\includegraphics[width=\columnwidth]{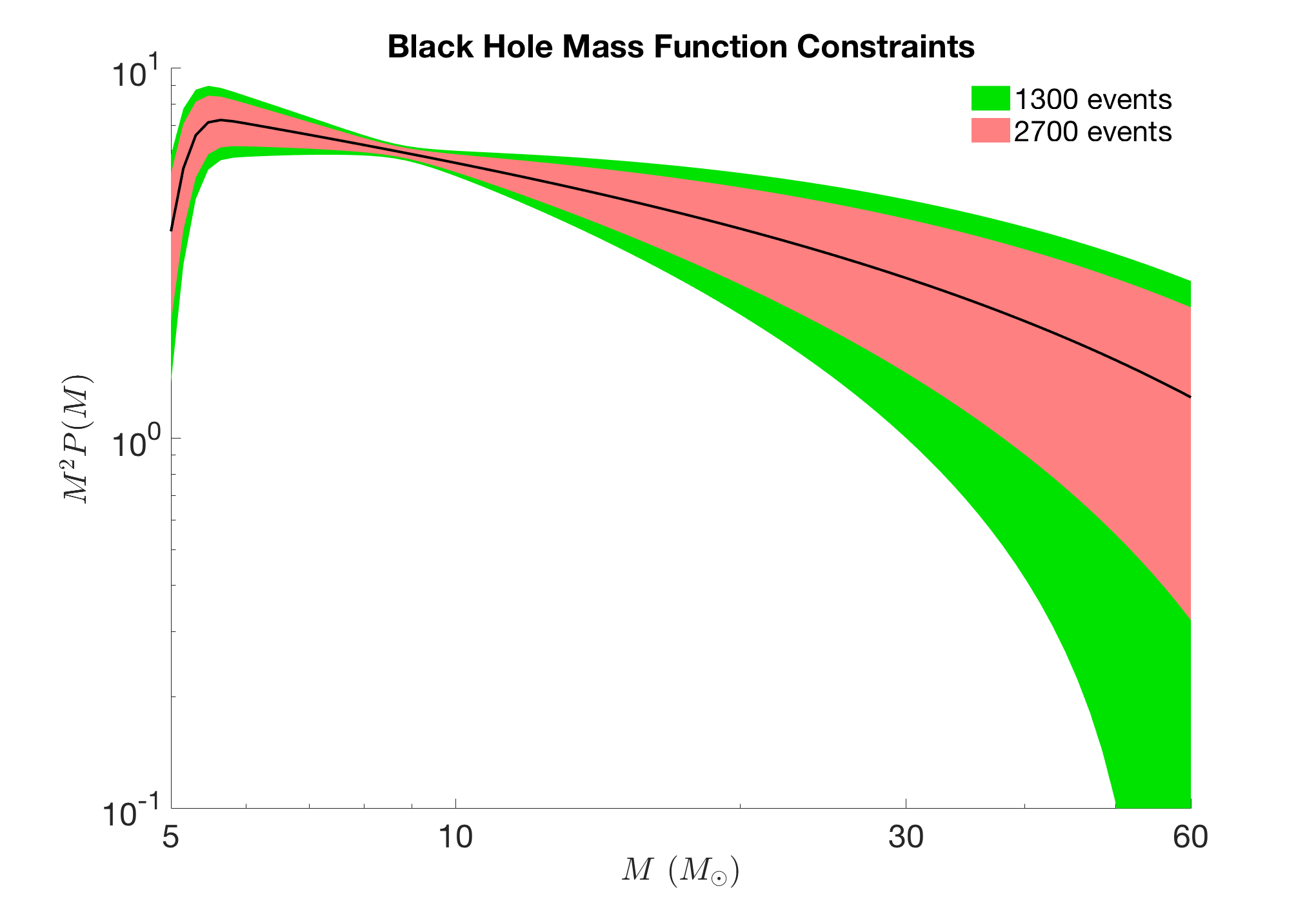}
\caption{Constraints on the BHMF, using 2D data.}
\label{fig:constrainedPm}
\end{figure}

In Fig.~\ref{fig:2Dvs1Dparams}, we compare the constraining power of the 
1D and 2D distributions on the mass ratio.  
We see that the 2D information provides more than an order-of-magnitude 
improvement in the determination of the mass-ratio parameter $\beta$. On the 
other hand, the 2D BH mass distribution affects the precision with which we 
can measure the mass slope $\alpha$ only at the 30$\%$ level, and has a 
negligible effect on the precision of $M_{\textrm{gap}}$ and $M_{\textrm{cap}}$.
This behavior should not come as a surprise. The parameter $\beta$ describes 
the PDF of the mass ratio $M_{2}/M_{1}$. When using the 1D BH mass distribution, 
we are ignoring the information on $M_{2}$ that is then reintroduced in the 2D case. 
Evidently, to probe the progenitors of the BBHs we need the information 
on both $M_{1}$ and $M_{2}$. 
The combined uncertainty on the BHMF is shown in Fig.~\ref{fig:constrainedPm}.

\begin{figure}[h!]
\includegraphics[width=0.95\columnwidth]{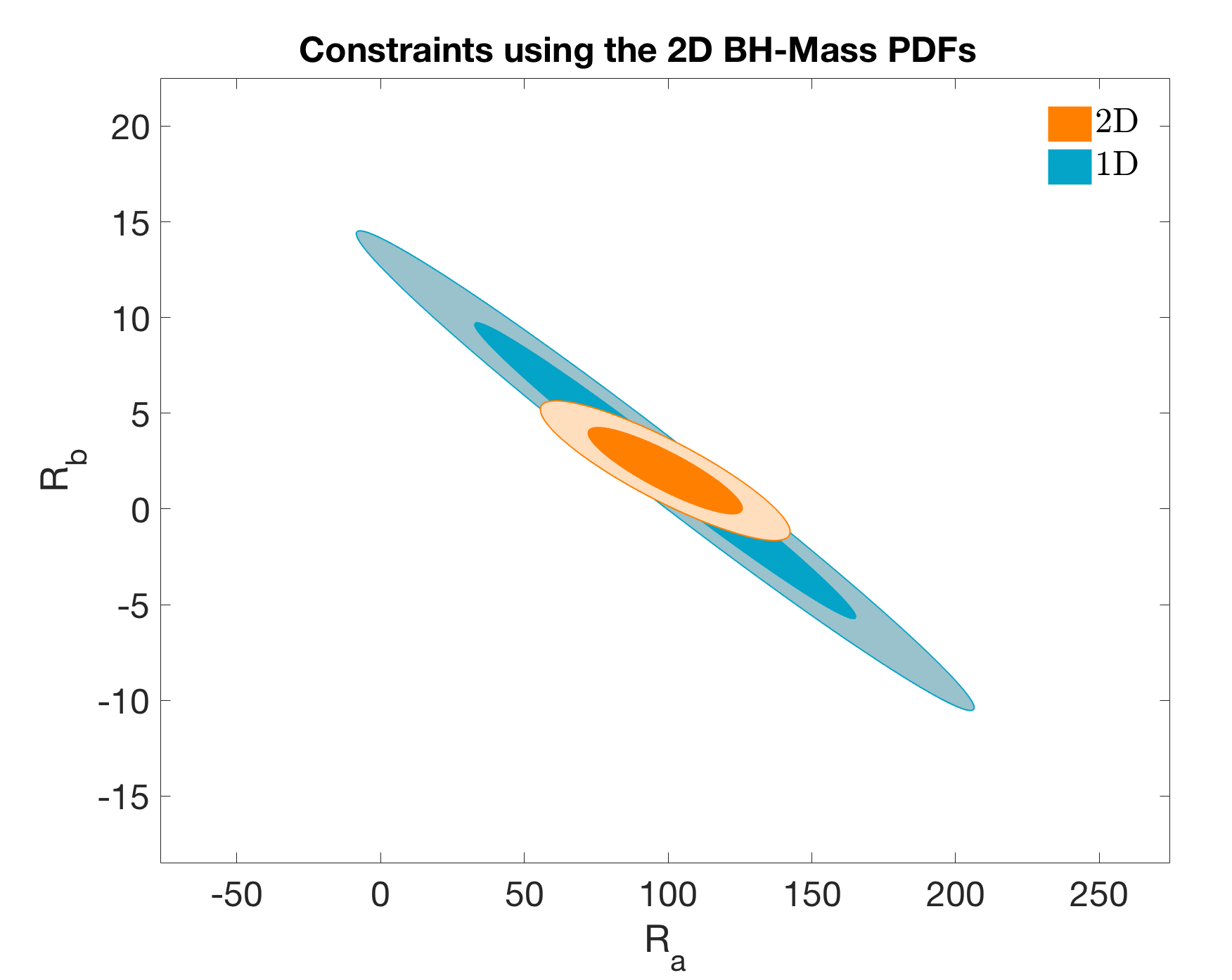}
\includegraphics[width=0.95\columnwidth]{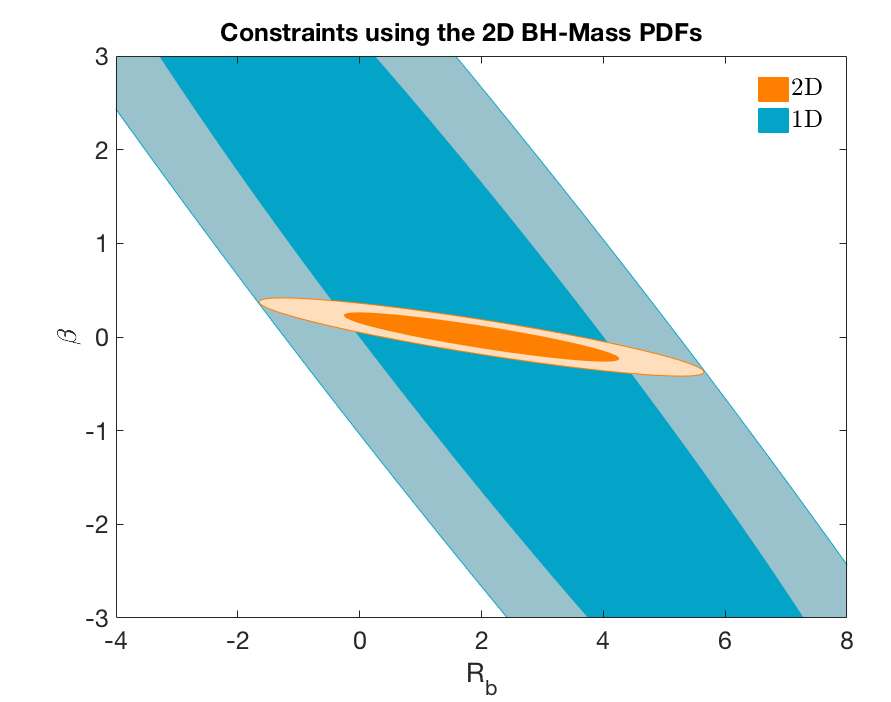}
\caption{{\it Top}: Constraints on the merger rate parameters, using 1D and 2D data 
(marginalizing over the other parameters. {\it Bottom}: The joint constraints for the 
mass-ratio power law and the merger-rate power law, showing the strong degeneracy,
which is then broken quite effectively when using the 2D information.}
\label{fig:betavsRb}
\end{figure}

We now return to the merger rate. It is clear that the mass ratio is degenerate 
with the rate of BBH mergers (a higher $\beta$ yields more observed events, as mergers 
between higher masses are easier to detect), but as we shall advocate, the 2D information 
can be used to break this degeneracy. Fig.~\ref{fig:betavsRb} demonstrates the  
degeneracy and also shows how using the full 2D information is efficient in breaking 
it and allowing for tighter constraints to be set on both the merger rate and mass ratio.   

Since advanced LIGO, similarly to other planned experiments, is expected to be 
iteratively improved throughout the coming decade, it is also interesting to examine 
how the constraints inferred from the data will incrementally improve with increased 
observation time. 
In Table~\ref{tab:Scenarios}, we present a full list of individual parameter constraints 
for a series of scenarios with respect to the observation time and imposed $S/N$
detection threshold. We argue that the choice of detection threshold is an important point 
to consider. In the initial stages of GW exploration, as long as the focus is on singular 
merger events, it is prudent to limit the number of false detections considerably. 
However, when striving to acquire a large statistical ensemble of events, it might 
be worthwhile to lower the threshold and accept more events, even at the 
price of allowing for a few spurious events to be included in the dataset. 
This will introduce more noise in the mass distribution measurements, but 
as long as this is washed out by the Poisson noise in each mass bin, relaxing 
the bound may be preferred.

\begin{figure}
\begin{centering}
\includegraphics[width=\columnwidth]{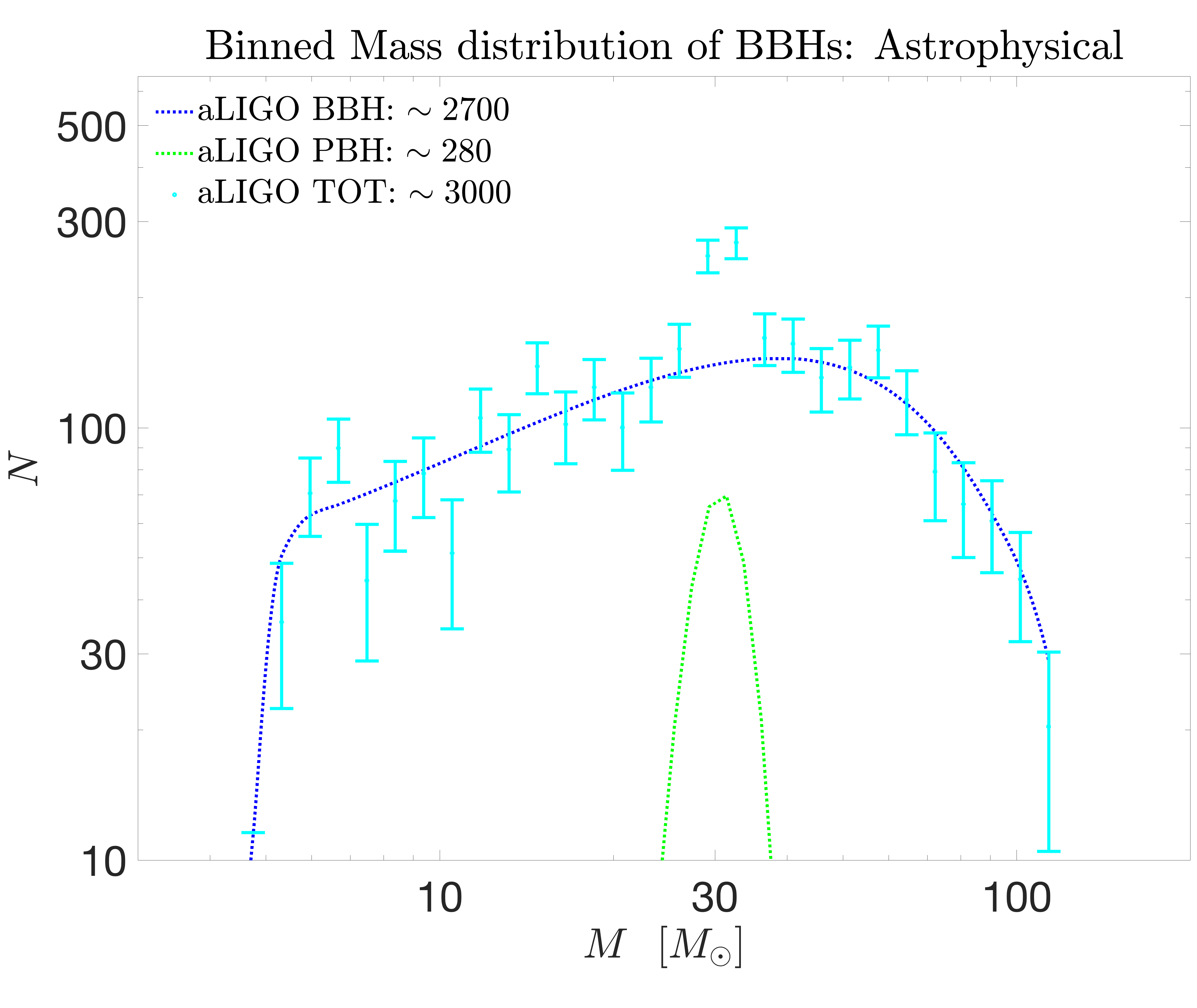}
\end{centering}
\caption{The logarithmically-binned distribution of the mass of the heavier 
component in BBH mergers, including mergers of both stellar black hole 
and primordial black holes (with a merger rate consistent with the assumption 
that they make up all of dark matter in the Universe, see Ref.~\cite{Bird:2016dcv}).}
\label{fig:PBH}
\end{figure}

Finally, we end this section with an example of how the merging mass statistics can
be used to probe particular models of binary progenitors. 
We focus on the primordial black hole model of \cite{Bird:2016dcv}, which was
mentioned above. In Fig.~\ref{fig:PBH}, we show a prediction for the mass
spectrum of all detected merger events with advanced LIGO at design sensitivity, 
counting both stellar black holes and primordial ones (assuming the abundance of 
the latter is such that they make up the dark matter in the Universe). For the purpose 
of this exercise, we take a Gaussian centered at $30\,M_\odot$ with a width of 
$3\,M_\odot$ for the PBH mass function (models of PBH formation generally 
do not have clear predictions for the mass function, while current experimental constraints 
limit much wider distributions). As can be seen, with several years of LIGO 
data, a several-$\sigma$ detection of PBHs can be made, or conversely a limit 
on the fraction of dark matter in PBHs can be inferred, assuming a particular form for
the PBH mass function (conventionally, one adopts a delta-function mass function 
when calculating constraints of this type). This is consistent with the rough estimate in 
Ref.~\cite{Bird:2016dcv}. While other methods have been proposed to constrain 
PBH dark matter in the stellar-mass range \cite{Mediavilla:2009um,Munoz:2016tmg,
Brandt:2016aco,Schutz:2016khr}, some even based on GW measurements 
\cite{Raccanelli:2016cud, Cholis:2016kqi}, we conclude that a simple examination of 
the mass spectrum is an efficient way of testing this scenario. We leave a more detailed 
investigation of this to future work \cite{kovetz}.

\begin{table*}
\setlength{\tabcolsep}{10pt}
\def\arraystretch{1.5}
    \begin{tabular}{|cccccccc|}
         \hline
         Scenario & $N$ &   $\sigma_\alpha$& $\sigma_{M_{\rm gap}} (M_{\odot})$ &  $\sigma_{M_{\rm cap}} (M_{\odot})$ & $\sigma_{R_a}$ (Gpc$^{-3}$yr$^{-1}$) & $\sigma_{R_b}$ & $\sigma_\beta$ \\
            \hline 
     $S/N>8$ and 1 years & 440   & 0.41 & 0.24 & 27.29 & 43.46 & 3.65 & 0.42\\
            \hline             
     $S/N>8$ and 3 years & 1330 &  0.23 & 0.14 & 15.75 & 25.09 & 2.11 & 0.24\\
            \hline 
     $S/N>8$ and 6 years & 2670 & 0.17 & 0.10 & 11.14 & 17.74 & 1.49 & 0.17\\
            \hline 
     $S/N>10/\sqrt{2}$ and 6 years & 3790 & 0.14 & 0.08 & 9.43 & 15.01 & 1.13& 0.14\\
            \hline 
     $S/N>8/\sqrt{2}$ and 6 years & 7050 & 0.11 & 0.06 & 7.87 & 11.85 & 0.72 & 0.11\\
            \hline 
            \end{tabular}
    \caption{Individual 1-$\sigma$ constraints on the BHMF, the merger rate and the 
    mass ratio parameters, under different scenarios. In addition to the standard criterion
    adopted by the LIGO collaboration of a signal-to-noise threshold {\it per detector} of 8, 
    we also consider thresholds of either 10 or 8 for two
    detectors combined. Lower thresholds yield larger statistical ensembles, obviously.
    For comparison, fiducial values in our analysis were taken to be $\alpha = 2.35$, 
    $M_{\rm gap} = 5 M_{\odot}$,   $M_{\rm cap} = 60 M_{\odot}$, $R_{a} = 99$ Gpc$^{-3}$yr$^{-1}$, 
    $R_{b} = 2$ and $\beta =0$. We treat 1 year of observation as 365 full days of data collection (duty cycle of unity).}
    \label{tab:Scenarios}
\end{table*}

\section{Discussion}
\label{sec:Discussion}

Our analysis includes a few caveats that should be pointed out.
Most evidently, a limitation of our results is that they rely on specific model choices, 
which include a (minimal) number of assumptions.  
For example, we assume a sharp cutoff at lower masses. This choice is motivated by 
both the currently available data (see Fig.~\ref{fig:Pm1Measured}) and by theoretical models of core-collapse 
supernovae whereby the instabilities driving the explosion have a rapid timescale ($<200\,{\rm ms}$), 
which have been shown to exhibit a NS-BH mass gap \cite{Belczynski:2011bn,Fryer:2011cx}. 
Nevertheless, it would be interesting to consider other choices, especially as more data 
is collected and the shape of the BHMF at low masses begins to unveil itself.

Another assumption we have made is that the BBH merger rate is 
mass-independent. While this assertion is supported in some models which 
take into account the explosion mechanism, the metallicity history and the time
delay distribution (see e.g.\ \cite{Fryer:2011cx, Dvorkin:2016wac}),
it presents a source of additional uncertainty. In follow up work, our analysis can be 
extended by incorporating the dependence of the merger rate on both mass 
and metallicity, and following the cosmic history of star formation and metallicity distribution, 
as well as the distribution of the delay time between formation and merger of the binaries. 
This can be done in the context of a galaxy evolution model, as carried out in 
Ref.~\cite{Dvorkin:2016okx}, for example. This model currently assumes that the two 
masses in each binary have independent distributions. It would be intriguing 
to generalize this method, accounting for different progenitor scenarios and 
incorporating the corresponding expectations for the mass ratio, and then proceed to 
investigate how well these models can be probed with future measurements.

We have focused on mass measurements in this work, neglecting the spin
of the black holes\footnote{In calculating Eq.~(\ref{eq:dN}), we set $\alpha_f$, the final 
spin parameter (which affects the result of $z_{\rm max}(M_1,M_2)$), to $0.67$ for all 
merger events. Relaxing this assumption, however, has a negligible effect.}.
There is definitely motivation to consider how well the distribution of initial spins of the merging 
black holes can be measured with future data, and make the connection with theoretical
predictions. We leave this for future work. 

We have also not included any discussion regarding systematic bias in
the parameter estimation of individual BBH coalescence events. In Ref.~\cite{Littenberg:2012uj},
it was shown that for events with SNR$<$50 (applicable for LIGO), any bias in characterizing
the GW events, introduced by the use of current waveforms, remains within the relevant 
statistical errors associated with the widths of the posterior distribution functions. More recently, 
in \cite{Pankow:2016udj}, it was also shown that even if the location on the sky and the distance 
to the binary are well known, either using an electromagnetic counterpart signal, or in the 
future by previous observations of the system \cite{Sesana:2016ljz} with eLISA \cite{AmaroSeoane:2012je},
the accuracy in measuring the spins and masses of the binary BHs does not improve significantly.

Care should be taken when directly comparing our results with forecasts made by the LIGO
collaboration and others. Beyond specific choices of parameters, which should not lead to 
any qualitative differences, we also ignore the fact that when searching for GW events in the 
LIGO data, the current bank of GW waveforms contains certain limitations, such as 
a total mass limit of $M_1+M_2<100\,M_\odot$ \cite{TheLIGOScientific:2016pea}. Such 
massive events, however, will yield very powerful signatures and may still be detected via 
the burst trigger or wavelet template searches. To facilitate the comparison with 
Ref.~\cite{TheLIGOScientific:2016pea}, we adopted the same value for the low mass cutoff. 
As also noted in Ref.~\cite{TheLIGOScientific:2016pea}, however, the total number of observed 
black holes in a given observation time will depend on this choice (the fixed fiducial merger 
rate amplitude we set means that a lower mass cutoff will result in a smaller abundance of 
more massive BHs, which are also more easily detectable).

Lastly, we advocated  that it may be worthwhile to lower the $S/N$ threshold 
imposed on the merger candidates identified by the template fitting process in order 
to enlarge the statistical ensemble used for inference of the BHMF and mass ratio 
parameters. Table I demonstrates the potential gain from a larger sample of BHs. 
However, it is important to emphasize that our analysis does not model the trigger 
events and does not account for any systematics or uncertainties that may be introduced 
by this process, which are beyond the scope of this paper.
 
\section{Conclusions}
\label{sec:Conclusions}

In this work we have investigated how measurements of gravitational waves from the merger 
of BBHs stand to advance our understanding of the mass distribution of BHs, thereby opening 
a  completely new avenue to study the most basic motif in astrophysics, namely the physics 
of stars, as well as provide us with valuable information about their cosmic history. In order to 
assess the power of a large statistical ensemble of mass measurements, it was important to 
adopt a well-motivated model for the BHMF, the BBH merger rate and the binary mass ratio, 
and to properly take into account important ingredients such as the instrumental noise and 
the mass measurement errors. Doing so, we attained a prediction for the number of detected 
events as a function of the mass of the constituent black holes.

An immediate conclusion this approach enabled, for example, 
is that events with masses in a range similar to those of the first event detected by advanced LIGO 
(and considerably more massive than the previously known BHs from X-ray observations), 
are in fact the most likely to be detected by this experiment (see Figure~\ref{fig:1Ddist}).

We then proceeded to study in detail how well the BHMF, the mass ratio, and the merger rate 
can be constrained with future data. We found that once LIGO reaches its design sensitivity, 
an expected number of $\gtrsim400$ black hole mergers will be observed per year, yielding 
remarkable constraints on the BHMF parameters. Notably, it will allow a measurement of the 
BHMF power law to better accuracy than our current best constraints on the stellar IMF power 
law in the heavier mass tail. By the time advanced LIGO finishes its planned full run, less than 
a decade from now, these constraints will more than double in accuracy. Together with the 
increasingly tight bound on the higher mass end of the BHMF, which probes the efficiency with 
which the heaviest stars maintain their mass, these results may therefore have more to say 
about the stellar mass distribution than the measurements of stars themselves(!).

The excellent expected sensitivity to the mass gap $M_{\rm gap}$, as indicated by our results, 
is such that if the true minimum mass of stellar black holes is indeed $\sim5\,M_\odot$, 
as indicated by (the very limited number of) current observations, then we can reject the hypothesis 
that the distribution extends all the way down to the upper limit on neutron star masses of 
$\simeq 2\,M_\odot$ at high significance. The uncertainty on the upper limit to 
stellar BH masses is expected to decrease to less than a decade in mass by the end of the 
LIGO run, providing an (indirect) observational handle on stars in their Wolf-Rayet phase. 

Another important conclusion of this work is that in order to exhaust the information from the 
mass measurements of merging black holes, it is imperative to focus on the two-dimensional 
mass distribution (of $M_{1}$ and $M_{2}$), shown in Fig.~\ref{fig:2Ddists}. This provides 
sensitivity to the progenitor models and breaks degeneracies between the merger rate 
and BHMF parameters and the binary mass ratio, as shown in Fig.~\ref{fig:2Dvs1Dparams}. 
We demonstrated that the effect of the merger rate parameters and the mass ratio on the observed 
mass distribution are tightly connected, and showed that the constraints on the mass ratio 
improve by more than an order of magnitude when using the full 2D information (see Fig.~\ref{fig:betavsRb}). 
When modeled as a power law, advanced LIGO should yield better than $10\%$ constraints on 
the value of $\beta$, its power law index (see Table~\ref{tab:Scenarios}).

These findings thus leave room for more detailed modeling of the BBH progenitor 
mechanisms---and the interplay between them---to be efficiently probed with the advent of thousands of 
GW detections from BH mergers. Future work will also consider next generation detectors such 
as the proposed Einstein Telescope \cite{Sathyaprakash:2012jk}, which can reach much greater 
sensitivities and would therefore probe the binary formation and merger history into higher redshifts, 
enabling tests of more extensive models than considered here. Finally, the more exotic 
example discussed above, of primordial black holes, provides a proof-of-concept that GW 
measurements can be used to derive constraints on the primordial spectrum of fluctuations 
which is imprinted onto the early Universe over cosmological scales by cosmological inflation 
\cite{Carr:2016drx,Munoz:2016pre}. This is another exciting topic to be studied in future work.

In conclusion, our results demonstrate that gravitational wave astronomy in the next decade will 
open up a novel and unique observational window on our Universe. If treated with the appropriate 
tools, as suggested above, this information can provide unprecedented insights 
into the physics of black hole formation, the survival of heavy stars, the history of star formation, 
the explosion mechanism of core-collapse supernovae, the nature of binary black hole progenitors, 
and even the existence of peaks in the primordial power spectrum of density fluctuations. 
Clearly, measurements of black holes will be enlightening. 

\acknowledgements
We thank Yacine Ali-Ha\"imoud, Bruce Allen, Simeon Bird, Irina Dvorkin, Julian Mu\~noz and 
Joe Silk for useful discussions. IC thanks the organizers of the GW $\&$ Cosmology workshop in DESY, Hamburg, Germany. 
This work was supported by NSF Grant No. 0244990, NASA NNX15AB18G and the Simons Foundation.

\end{document}